\documentclass[superscriptaddress,amsmath,amssymb,aps,prb,twocolumn,floatfix]{revtex4-2}

\usepackage{graphicx}
\usepackage{bm}
\usepackage[colorlinks, linkcolor=mycol, citecolor=mycol, urlcolor=mycol, breaklinks]{hyperref}
\usepackage{braket}
\usepackage{bbold}
\usepackage{bm}
\usepackage{xcolor}
\usepackage{comment}

\newcommand{\mi}{ {\rm i} }
\newcommand{\me}{ {\rm e} }
\newcommand{\id}{\mathbb{1}}
\newcommand{\bi}{ {\bm{i}} }
\newcommand{\bj}{ {\bm{j}} }
\newcommand{\bq}{ {\bm{q}} }

\definecolor{mycol}{RGB}{10,55,130}

\begin{document}
\title{Dark state semilocalization of quantum emitters in a cavity} 

\author{T. Botzung}
\affiliation{Universit\'{e} de Strasbourg and CNRS, ISIS (UMR 7006) and icFRC, 67000 Strasbourg, France}
\affiliation{Institute for Quantum Information, RWTH Aachen University, D-52056 Aachen, Germany
Peter Grünberg Institute, Theoretical Nanoelectronics, Forschungszentrum Jülich, D-52425 Jülich, Germany}

\author{D.~Hagenm\"uller}
\affiliation{Universit\'{e} de Strasbourg and CNRS, ISIS (UMR 7006) and icFRC, 67000 Strasbourg, France}

\author{S.~Sch\"utz}
\affiliation{Universit\'{e} de Strasbourg and CNRS, ISIS (UMR 7006) and icFRC, 67000 Strasbourg, France}
\affiliation{IPCMS (UMR 7504), CNRS, 67000 Strasbourg, France}

\author{J.~Dubail}
\affiliation{Universit\'{e} de Lorraine, CNRS, LPCT, F-54000 Nancy, France}
\affiliation{Universit\'{e} de Strasbourg and CNRS, ISIS (UMR 7006) and icFRC, 67000 Strasbourg, France}

\author{G.~Pupillo}
\thanks{pupillo@unistra.fr}
\affiliation{Universit\'{e} de Strasbourg and CNRS, ISIS (UMR 7006) and icFRC, 67000 Strasbourg, France}
\affiliation{Institut Universitaire de France (IUF), 75000 Paris, France}

\author{J.~Schachenmayer}
\thanks{schachenmayer@unistra.fr}
\affiliation{Universit\'{e} de Strasbourg and CNRS, ISIS (UMR 7006) and icFRC, 67000 Strasbourg, France}
\affiliation{IPCMS (UMR 7504), CNRS, 67000 Strasbourg, France}

\date{\today}

\begin{abstract}
We study a disordered ensemble of quantum emitters collectively coupled to a lossless cavity mode. The latter is found to modify the localization properties of the ``dark'' eigenstates, which exhibit a character of being localized on multiple, noncontiguous sites. We denote such states as semilocalized and characterize them by means of standard localization measures. We show that those states can very efficiently contribute to coherent energy transport. Our paper underlines the important role of dark states in systems with strong light-matter coupling.
\end{abstract}

\maketitle

\section{Introduction}

When quantum emitters and a cavity mode coherently exchange energy at a rate faster than their decay, hybrid light-matter states play an important role~\cite{Tavis_Exact_1968,Kimble_Strong_1998,Raimond_Manipu_2001}. Such polaritonic states are superpositions composed of ``bright'' emitter modes and cavity photons, while numerous remaining emitter states have no photon contribution, i.e.~remain ``dark''. Collective strong light-matter coupling has been intensively pursued in atomic~\cite{Haroche_1983,Carmichael_1989,Thompson_Observ_1992} and condensed matter physics~\cite{Weisbuch_1992,imamoglu1996,Lidzey,Wallraff_2009,Esteve_2010,Tabuchi_2014}. Very recently, strong collective coupling has been explored as tool to engineer fundamental properties of matter, e.g.~the critical temperature of superconductors~\cite{Sentef_2018,Thomas_Explor_2019} or chemical reaction rates~\cite{Thomas_Tilting_2018,Kena-Cohen_Polariton_2019,Lather_Cavity_2019,Thomas_Ground_2016,Hutchison_Modify_2012,Herrera_2016,galego_2016,Flick_2017}. Much interest is currently raised by the possibility of modifying energy~\cite{Coles2014,Feist_Extrao_2015,Schachenmayer_Cavity_2015,Zhong_2017,Lerario2017,Reitz_Energy_2018,Du2018,Schafer2019} and charge~\cite{Orgiu_Conduc_2015,Hagenmuller_Cavity_2017,Hagenmuller_Cavity_2018,Schafer2019} transport.

For transport, disorder plays a crucial role. It is well studied that coherent transport is inhibited due to Anderson localization (AL)~\cite{Evers_Anders_2008}. Here, an arbitrarily small disorder can lead to a localization of eigenstates in 1D and 2D~\cite{Anderson_Absenc_1958,Evers_Anders_2008}, while in 3D a metal--insulator transition driven by the disorder strength occurs~\cite{Abrahams_1979,Evers_Anders_2008}. Here, we study the fate of this phenomenon in a cavity.  It is known that polaritonic states are largely unaffected by disorder~\cite{Houdre_Vacuum_1996} and the impact of disorder on polariton physics for laser-driven setups has been extensively explored~\cite{Eastham_Bose_2001, Litinskaia_Inhom_2001, Marchetti_Therm_2006, Marchetti_Absor_2007, Kirton_Intro_2019}. While for transport problems it is known that polariton states can lead to a considerable enhancement of energy transmission~\cite{Feist_Extrao_2015,Schachenmayer_Cavity_2015,Zhong_2017,Du2018,Lerario2017,Reitz_Energy_2018,Schafer2019}, the localization and transport properties of the dark states have remained largely unexplored. It is clear that disorder leads to a mixing of the bright with the dark states~\cite{Gonzalez-Ballestero_Uncoup_2016}, which alters the usual description of light-matter coupling. Addressing these issues is for example important for applications of radiative energy transmission in mesoscopic systems.

\begin{figure}[t]
    \centering
    \includegraphics[width=1\columnwidth]{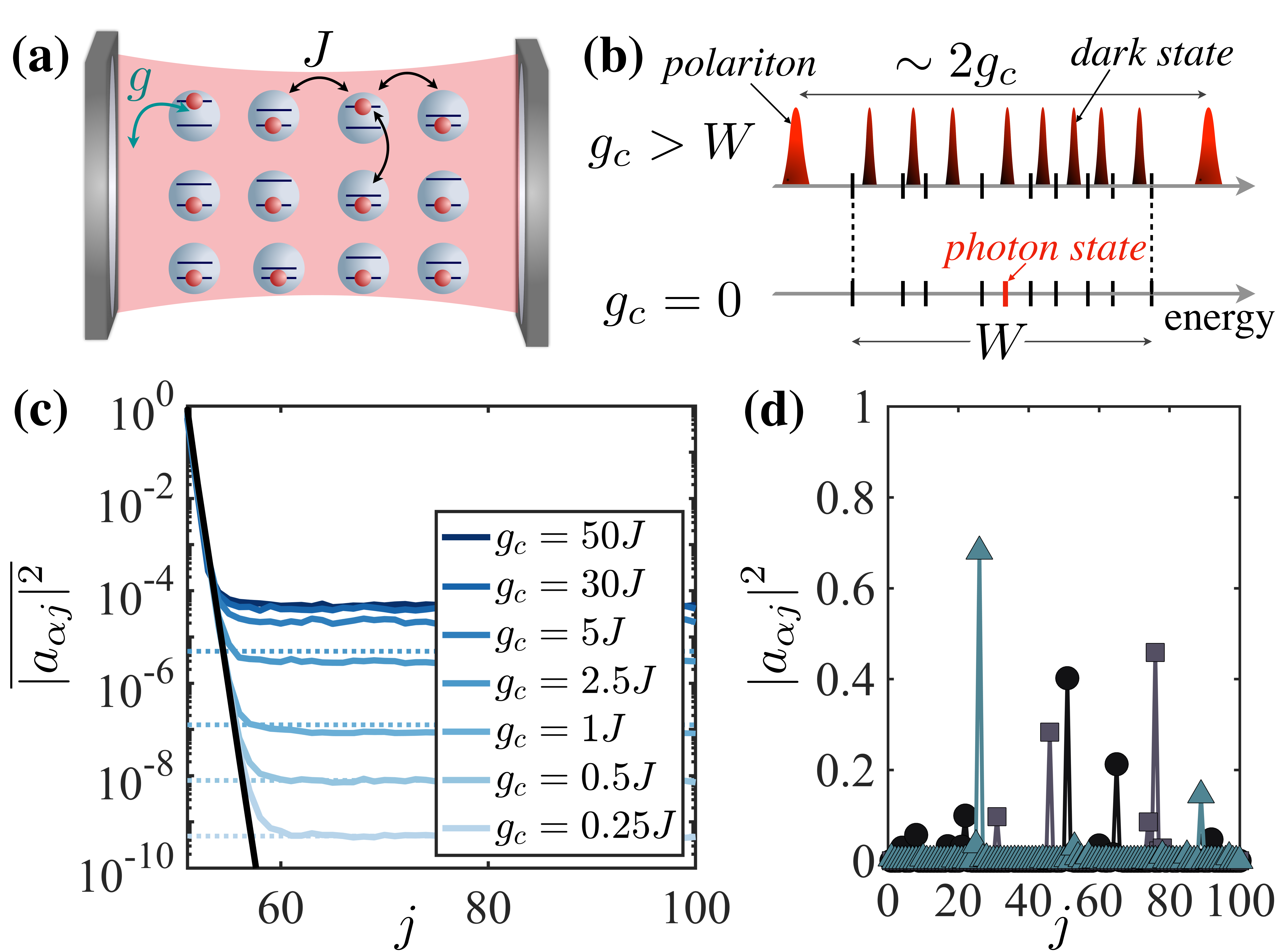}
    \caption{\textbf{(a)} An excitation can hop with rate $J$ on a disordered 3D lattice with $N$ sites. Local transitions are coupled to a cavity with collective strength $g_{c}\equiv g \sqrt{N}$. \textbf{(b)} For $g_{c}=J=0$, the $N$ bare levels are randomly distributed in $[-W/2,W/2]$. For $g_{c} > W$, the spectrum contains two polariton eigenstates (splitting $\sim 2 g_{c}$) and $N-1$ dark states lying in between the bare levels. \textbf{(c)} Modification of the disorder-averaged weights of a dark eigenstate localized in the middle of a chain (1D for convenience) with $N=100$, $2000$ realizations, $W=25 J$. In addition to exponential localization at short distances (black line, $g_c=0$), a constant tail appears for $g_c>0$ (dashed lines: perturbative results). \textbf{(d)} Single disorder realization: Three dark eigenstates are shown for $g_{c}=50 J$ and $W=25 J$.}
    \label{fig:fig1}
\end{figure}

In this paper, we investigate a simple model for AL and coherent energy transport with $N$ emitters collectively coupled to a cavity mode [Fig.~\ref{fig:fig1}\textbf{(a)}]. We focus on the impact of the cavity coupling on localized eigenstates, i.e.~for a disorder strength much larger than the excitation hopping rate. We focus on dark states and find that they exhibit several surprising features: for any strength of light-matter interactions, they acquire a squared amplitude $\sim 1/{N}$, on average, for arbitrary distances.  While their photon weight vanishes, they can remain localized according to standard localization measures, such as the inverse participation ratio (IPR) in the thermodynamic limit. However, we find that localization is distributed over multiple sites, which can be arbitrarily distant from one another. For large cavity coupling, they can be considered as hybridizations of a few localized states of the uncoupled system, and their energy lies in between those of the bare states. This results in semi-Poissonian statistics of energy level spacings, which neither corresponds to a fully localized nor extended phase. We find that semilocalized states are responsible for diffusive-like dynamics, which is at odds with their localized nature.  On average, the exponential decay of an excitation current with $N$ for AL can be turned into an algebraic decay $\sim 1/N$, and can thus dominate over the $\sim 1/N^2$ contribution expected from polariton states. The non-local nature of the cavity-coupling makes these effects independent of the dimensionality.

Our results are directly relevant for  setups with condensed matter interacting with confined electromagnetic fields (both close to a vacuum state~\cite{Sentef_2018,Thomas_Explor_2019,Thomas_Tilting_2018,Kena-Cohen_Polariton_2019,Lather_Cavity_2019,Thomas_Ground_2016,Hutchison_Modify_2012,Herrera_2016,galego_2016,Flick_2017,Coles2014,Feist_Extrao_2015,Schachenmayer_Cavity_2015,Zhong_2017,Lerario2017,Reitz_Energy_2018,Du2018,Schafer2019,Orgiu_Conduc_2015,Hagenmuller_Cavity_2017,Hagenmuller_Cavity_2018,Schafer2019} or laser-driven ~\cite{carusotto_2013,Torma_2014,Sanvitto_Road_2016,Putz_Spect_2016,Deng2002,Kasprzak2006,Balili2007,Amo2009,Amo2010,Keeling2010,Deng_Excito_2010,Juggins_Coher_2018}), in particular for recent coherent transport setups~\cite{Cadiz_2018,Wang_2019}. Furthermore, in the past years, localization has been extensively experimentally explored with controlled disorder in cold atom systems~\cite{Billy_Direc_2008,Roati_Ander_2008,Kondov_Three_2011,Jendrzejewski_Three_2012}, even in interacting many-body regimes (many-body localization)~\cite{Schreiber_Obser_2015}. Specifically, experiments using Dicke model realizations based on Raman dressed hyperfine ground-states of intra-cavity trapped atoms, as recently achieved~\cite{Dimer_Propo_2007, Zhang_Dicke_2018}, could be used to study the semilocalized physics described here. We propose a specific cold atom implementation of our model in Appendix~\ref{sec:app:AMO}.  Models of emitters interacting with a single cavity mode are also formally similar to central spin models~\cite{gaudin1976diagonalisation,dukelsky2004colloquium} that have been very successful in modeling hyperfine interactions of quantum dots surrounded by a bath of nuclear spins~\cite{schliemann2003electron,bortz2007exact,faribault2013integrability}; the effects of disorder in central spin models have been investigated recently in~\cite{hetterich2017noninteracting,hetterich2018detection} in connection with many-body localization.

The remainder of the paper is organized as follows. In Sec.~\ref{sec1}, we introduce the model under consideration. In Sec.~\ref{sec2}, we discuss the physics of the model in various regimes and show that localized eigenstates acquire an average probability amplitude for arbitrary distances for strong enough light-matter couplings. This ``semilocalization'' behavior is characterized by standard localization measures such as the return probability, the inverse participation ratio (Sec.~\ref{subsec21}), and the level statistics (Sec.~\ref{subsec22}). In Sec.~\ref{sec3}, we explore the model dynamics by computing the excitation current flowing through the system and the time-dependence of the mean square displacement.  We provide a conclusion in Sec.~\ref{sec5}.

\section{Model}
\label{sec1}

We start by considering a 3D cubic lattice of $N$ two-level systems embedded in a cavity. The Hamiltonian ($\hbar \equiv 1$) is $\hat{H}=\hat{H}_{0}+\hat{H}_{\rm I}$, with 
\begin{align}
    \hat{H}_{0}=\omega_{c}\hat{a}^\dag \hat{a}+ \sum_{\bi} (\omega_{e}+w_{\bm i}) \hat \sigma_{\bi}^{+}\hat \sigma_{\bi}^{-} -J \sum_{\langle \bi,\bj \rangle} \hat \sigma_{\bi}^{+} \hat \sigma_{\bj}^{-},
\end{align} and 
\begin{align}
    \hat H_{\rm I} =  g \sum_{\bi} (\hat a \hat \sigma_{\bi}^{+} + \hat a^\dag \hat \sigma_{\bi}^{-} ).
\end{align}
We restrict our discussion to a Hilbert space with a single excitation, i.e.~$\sum_\bi \hat \sigma_\bi^{+} \hat \sigma_{\bi}^{-} + \hat a^\dag \hat a = \id$. Then, there are $N+1$ basis states, $\ket{\bi,0}$, $\ket{G,1}$, denoting states with an excitation on site $\bi$, or in the cavity, respectively. Also considering the state without excitation, $\ket{G,0}$, the spin lowering and photon annihilation operators are defined as $\hat \sigma_{\bi}^{-}=\ket{G,0}\bra{\bi,0}$ and $\hat a=\ket{G,0}\bra{G,1}$. In all numerical calculations, we consider the cavity mode (frequency $\omega_c$) in resonance with the average emitter transition ($\omega_e$), i.e.~$\delta\equiv \omega_{e}-\omega_{c}=0$. The third term in $\hat H_{0}$ governs hopping (rate $J$) between nearest neighbor sites, indicated by the notation $\langle\bi, \bj \rangle$. Assuming periodic boundaries, this term is diagonalized by introducing the operators $\hat b_{\bq} =\sum_{\bi} \exp(-\mi \bq \cdot \bi) \hat \sigma_{\bi}^{-}/\sqrt{N}$. The second term contains on-site disorder, with $w_\bi$ random variables uniformly distributed in $[-W/2,W/2]$. The other term $H_{\rm I}$ describes the Tavis-Cummings emitter-cavity coupling~\cite{Tavis_Exact_1968} with local strengths $g$. This term can be written in the form 
\begin{align}
H_{\rm I}=g_c (\hat a \hat b_{{\bm 0}}^\dag + \hat a^\dag \hat b_{{\bm 0}})
\end{align}
with the collective strength $g_c = g\sqrt{N}$, and couples the symmetric bright mode $\hat b_{\bq = 0}$ to cavity photons. Importantly, $g$ decreases with the cavity-mode volume $V$ as $g\sim 1/\sqrt{V}$~\cite{scully_zubairy_1997} and $g_{c}$ thus remains independent of $N$ for fixed density $N/V$. 

\section{Semilocalization}
\label{sec2}

\begin{figure*}[t]
    \centering
    \includegraphics[width=1\textwidth]{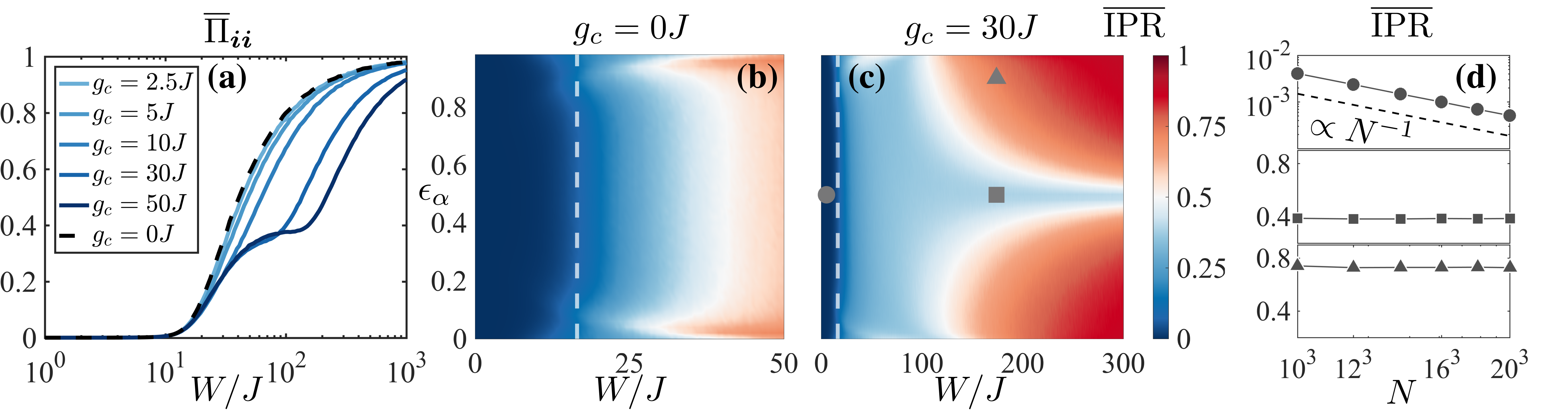}
    \caption{\textbf{(a)} Disorder-averaged return probability $\overline{\Pi}_{\bi \bi}$  as a function of $W/J$ (for the central site $\bi$ of a $N = 15^3$ cube, mean emitter splitting on resonance with cavity, $\delta=0$). For strong-couplings $g_c > W > W_{c}$, a plateau ($\overline{\Pi}_{\bi \bi}\simeq 0.4$) indicates a ``semilocalized'' regime. \textbf{(b-c)} Disorder-averaged inverse participation ratio $\overline{\mathrm{IPR}}(\epsilon_{\alpha})$ as a function of $W/J$ and the renormalized dark state energy $\epsilon_{\alpha}$ (bins of widths $0.02$, $\sim 100$ realizations, white dashed line: $W=W_{c}$). \textbf{(b)} $g_c = 0$ (no cavity); \textbf{(c)} $g_c = 30J$ (larger $W/J$-scale), showing an extended area with $\overline{\mathrm{IPR}}(\epsilon_{\alpha})\simeq 0.4$. \textbf{(d)} Finite-size scaling of $\overline{\mathrm{IPR}}(\epsilon_{\alpha})$ for the parameters corresponding to the symbols in \textbf{(c)}. Circle ($W=5J$, $\epsilon_{\alpha}=0.5$); square ($W=175J$, $\epsilon_{\alpha}=0.5$); triangle ($W=175J$, $\epsilon_{\alpha}=0.9$).}
    \label{fig:fig2}
\end{figure*}

\subsection{Semilocalized eigenstates}
\label{subsec21}

In the absence of disorder ($W=0$), $\hat H$ has two polariton eigenstates $\ket{\psi_\pm} = (\hat b_{\bm{q}=\bm{0}}^\dag \pm \hat a^\dag)/\sqrt{2} \ket{G,0}$ with energies $E_{\pm}=\pm g_c$, as well as $N-1$ uncoupled dark states $\ket{\psi_{\alpha\neq \pm}}=\hat b^\dag_{\bq \neq \bm{0}}\ket{G,0}$ with vanishing photon weight, $\langle G,1 \vert \psi_{\alpha\neq \pm} \rangle=0$. It is obvious that finite disorder ($W\neq 0$) leads to a coupling between the bright and the dark states since $\hat H_{\rm I}$ is non-diagonal in quasi-momentum space. The dark eigenstates therefore acquire a small photonic weight $\vert\braket{G,1|\psi_{\alpha\neq \pm}}\vert^{2} \sim 1/N$ (see Appendix \ref{sec:app:pert}), and become ``grey''. In the following we are interested in the modification of the emitter part of the system, and define the normalized emitter amplitudes as $a_{\alpha\bj} \equiv \langle \bj,0 \vert \psi_{\alpha} \rangle/\sqrt{\mathcal{N_\alpha}}$ with $\mathcal{N_\alpha} = 1-|\braket{G,1|\psi_\alpha}|^2$.

For $g_{c}=0$, $\hat H$ corresponds to a usual AL model, displaying a $W$-dependent mobility edge that determines a metal-insulator transition at $W_{c} \simeq 16.5 J$ (for energy states in the middle of the band)~\cite{Kramer_1981,Hofstetter_1993,Shklovskii_1993,Zharekeshev_1995}. While for $W \ll W_{c}$ the eigenstates $\ket{\psi_\alpha}$ resemble extended Bloch states, they are localized around given sites for $W>W_{c}$, e.g.~$\vert a_{\alpha\bj} \vert^{2}\propto \me^{-\vert \bi-\bj\vert/\xi}$ for a state localized on site $\bi$, with $\xi$ a $W/J$-dependent localization length. In the following, we investigate the case $g_{c},W\neq 0$, and focus on spectral and transport properties of the Anderson insulator for strong collective light-matter couplings $g_{c}>W>W_{c}$. 

The modification of AL in a cavity can be understood by first considering the eigenstates of $\hat H$ for $J=0$, in which case the spatial dimensionality becomes irrelevant. 
In second-order perturbation theory (see Appendix \ref{sec:app:pert}), a trivially localized eigenstate on site $\bi$, $\ket{\bi,0}$, for $g_{c}=0$ acquires an amplitude on site $\bj\neq \bi$ via the cavity, 
\begin{align}
b_{\bi\neq\bj} = \frac{g^2}{(w_{\bi}-w_{\bj})(w_{\bi} + \delta)},
\end{align}
valid for configurations with $g^2 \ll |(w_{\bi}-w_{\bj}) (w_{\bi}+\delta)|$. A lower bound for the squared amplitude of perturbed localized states is thus $|b_{\bi\neq\bj}|^2 \geq 4 g_c^4/(N^2 W^4)$, setting $\delta = 0$. In Appendix \ref{sec:app:pert} we use perturbation theory to derive that the averaged value over disorder realizations is 
\begin{align}
\overline{|b_{\bi\neq\bj}|^2} =\frac{4g_{c}^4\left[4-2\log(4)\right]}{N W^{4}}
\end{align}
 for large $N$. In Fig.~\ref{fig:fig1}\textbf{(c)} we show numerically that also for finite $J\ll W$ the weights of an eigenstate localized in the center of a 1D chain, logarithmically averaged over disorder realizations, maintains an exponentially localized profile at short distances, followed by a constant tail rising with $g_c$. The tails are consistent with our perturbative result for small $g$ (dashed lines) and saturate for strong couplings $g_c > W >J$. Note that a similar behavior was reported for dissipative couplings to a common reservoir~\cite{Celardo_2013, Biella_Subradiant_2013}.

For strong coupling ($g_c > W > J$), two polaritonic states $\ket{\psi_{\pm}}$ with $\vert\langle G,1\vert \psi_{\pm} \rangle \vert^{2} \approx 0.5$ and separated by a splitting $\sim 2 g_{c}$ (only slightly modified by disorder) emerge from the  band of width $W$. We find that the energies of the $N-1$ dark states  lie in between the $N$ bare ($g_c=0$) levels [Fig.~\ref{fig:fig1}\textbf{(b)}], which can be seen as a simple consequence of the ``arrowhead'' matrix shape of the single-excitation Hamiltonian for $J=0$, as we detail in Appendix~\ref{sec:app:arrow}. The strong cavity coupling leads to a hybridization between bare levels, close in energy, but not necessarily in real space. For a single disorder realization, the dark states appear strongly localized at multiple sites [Fig.~\ref{fig:fig1}\textbf{(d)}]. We term this behavior as \textit{``semilocalization''}.

Information about the spatial localization of dark eigenstates with energy $E_{\alpha}$ is given by the inverse participation ratio (IPR),
\begin{align}
\mathrm{IPR}(E_\alpha) = \sum_{i=1}^{N} {\vert a_{\alpha\bi}\vert^{4}}.
\end{align}
A finite, size-independent IPR indicates a localized eigenstate, while an IPR scaling as $1/N \to 0$ indicates an extended one. Initializing the system in the state $\ket{\bi,0}$, the infinite-time-averaged probability to find an excitation at site $\bj$ is 
\begin{align}
\Pi_{\bi\bj}=\lim_{T\to \infty} \frac{1}{T}\int_{0}^{T} dt P_{\bi\bj} (t),
\end{align}
 with $P_{\bi\bj} (t)\equiv \vert\langle \bj,0 \vert \phi (t) \rangle \vert^{2}$ and $\ket{\phi (t)}=\me^{-\mi\hat{H}t}\ket{\bi,0}$. The IPR is connected to the return probability $\Pi_{\bi\bi}$ by $\sum_{\bi} \Pi_{\bi \bi}  =  \sum_{\alpha} {\rm IPR} (E_{\alpha}) \mathcal{N}^2_\alpha$. The ${\rm IPR}(E_\alpha)$ can thus be interpreted as the contribution of a given eigenstate to $\sum_{\bi} \Pi_{\bi \bi}$. 

\smallskip

In Fig.~\ref{fig:fig2}\textbf{(a)}, we compute numerically the disorder average of ${\Pi}_{\bi \bi}$, $\overline{\Pi}_{\bi \bi}$, for the central site of a cubic lattice ($N = 15^3$). For $g_{c}=0$ (dashed line), $\overline{\Pi}_{\bi \bi}$ increases from $0$ (extended phase) to $1$ (localized phase) upon increasing the disorder strength $W/J$. Remarkably, we find that $\overline{\Pi}_{\bi \bi}$ exhibits a plateau $\simeq 0.4$ for $g_c > W >J$, which persists up to  large disorder strengths ($W \sim 100 J$ for $g_c=50J$). 

The disorder-averaged IPR, $\overline{\mathrm{IPR}}(E_\alpha)$, is shown in Figs.~\ref{fig:fig2}\textbf{(b-c)} as a function of $W/J$ for the AL model ($g_c = 0$) and for $g_c = 30J$. As we only focus on dark states (in the band of width $W$), we use a dimensionless, renormalized energy scale $\epsilon_{\alpha}= (E_{\alpha} - W/2)/W$ with $\epsilon_{\alpha}\in [0,1]$. For each disorder realization, we bin different levels into groups with equal energy width and then average over realizations in each bin. Figure~\ref{fig:fig2}\textbf{(b)} shows the emergence of localized states upon increasing $W/J$, starting from the edges of the spectrum. A strong cavity coupling [Fig.~\ref{fig:fig2}\textbf{(c)}] leads to \textit{three} distinct regimes: i) a delocalized region [$\overline{{\rm IPR}}(\epsilon_\alpha) \sim 0$ for $W\lesssim W_c$]; ii) a fully localized region [$\overline{{\rm IPR}}(\epsilon_\alpha) \sim 1$ for $W > g_c$]; and iii) an extended area with $\overline{{\rm IPR}}(\epsilon_\alpha) \sim 0.4$ where the dark states feature semilocalized characteristics consistent with the return probability and the results shown in Fig.~\ref{fig:fig1}\textbf{(c)}. The persistence of semilocalized states in the vicinity of $\epsilon_{\alpha}= 0.5$ ($\delta=0$) can be understood from the failure of perturbation theory, even for $W \gg g_c$. The energy separation between the two levels ($\bi_0$, $\bj_0$) closest to $\epsilon_\alpha=0.5$ is $(w_{\bi_0} - w_{\bj_0}) \sim W/N$. For them, the perturbation condition $g_c^2 \ll W (w_{\bi_0/\bj_0}+\delta)$ is violated for all $W$ considered in Fig.~\ref{fig:fig2}\textbf{(c)}, as they hybridize via the cavity. 

In Fig.~\ref{fig:fig2}\textbf{(d)}, we analyze the finite size scaling of $\overline{\mathrm{IPR}}(\epsilon_\alpha)$ in the three regions [for parameters corresponding to the symbols in Fig.~\ref{fig:fig2}\textbf{(c)}]. We observe that the IPR of semilocalized states does not scale with the system size. These states exhibit the same behavior as in the fully localized region, only with a reduced value, which is consistent with states localized on multiple sites. In contrast, $\overline{\mathrm{IPR}}(\epsilon_\alpha) \propto 1/N$ for extended states.

\subsection{Level statistics}
\label{subsec22}

\begin{figure}[tb]
    \centering
    \includegraphics[width=1\columnwidth]{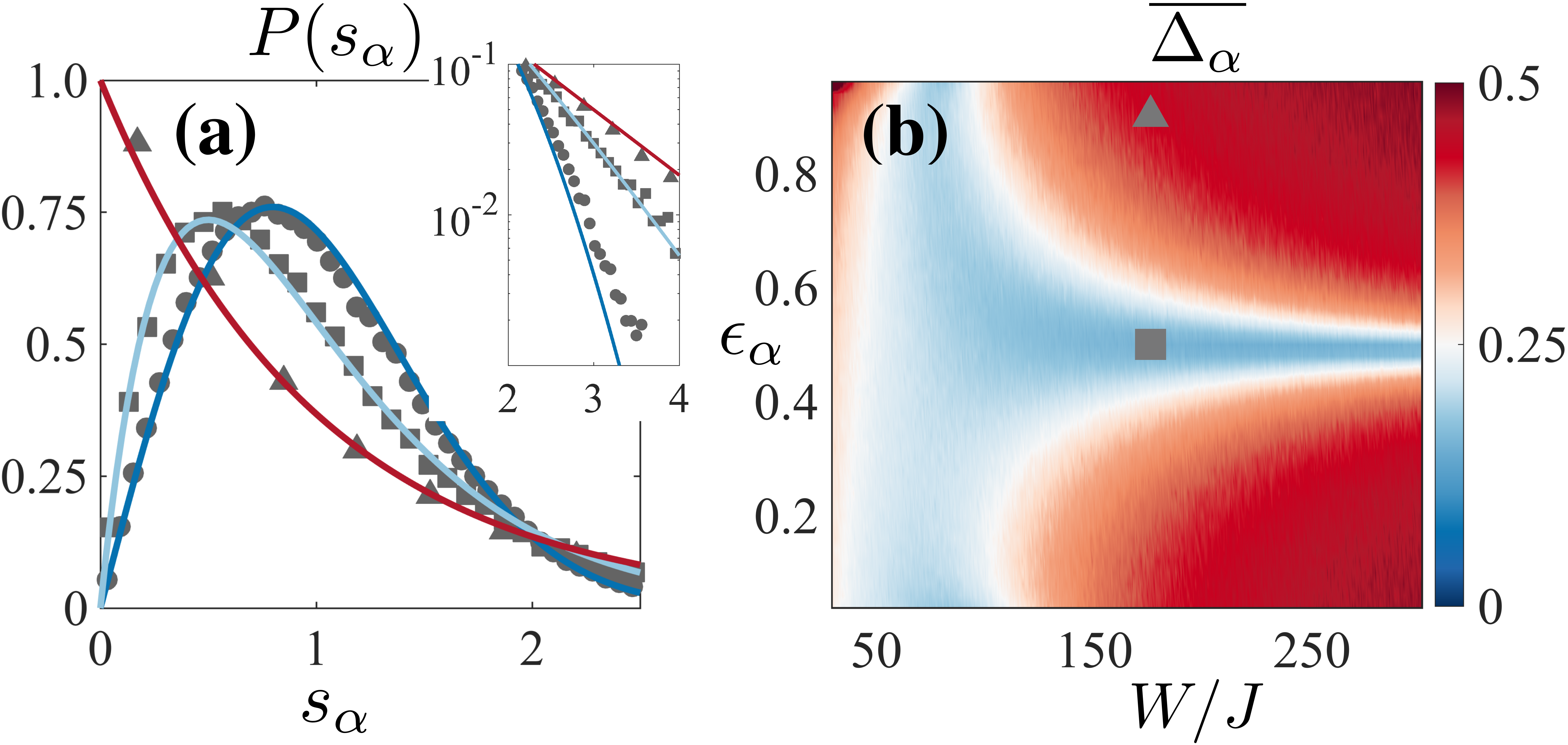}
    \caption{\textbf{(a)} Comparison of numerically computed distributions $P(s_{\alpha})$ (symbols, parameters as in~\ref{fig:fig2}\textbf{(c)}) with analytical formulas (lines, see text). Wigner-Dyson distribution (dark blue); Poissonian distribution (red); Semi-Poissonian distribution (light blue). Inset: tails of the distributions on a logarithmic scale. \textbf{(b)} Numerically computed ``dark state deviations'' $\overline{\Delta}_{\alpha}$ (see text) as a function of $W/J$ for $g_c=30J$.}
    \label{fig:fig3}
\end{figure}

Localization properties of eigenstates are also characterized by their level statistics~\cite{Zharekeshev_1995}. Here, we numerically analyze the probability distribution function $P(s_{\alpha})$ for spacings between adjacent eigenenergies, $s_{\alpha} = \epsilon_{\alpha+1} - \epsilon_{\alpha}$. In Fig.~\ref{fig:fig3}\textbf{(a)}, we plot $P(s_{\alpha})$ for eigenstates corresponding to the symbols in Fig.~\ref{fig:fig2}\textbf{(c)}. While in the delocalized region ($W \lesssim W_c$) $P(s_{\alpha})= \frac{\pi}{2} s_{\alpha} \exp( - \frac{\pi}{4} s_{\alpha}^{2})$ follows a Wigner-Dyson distribution, the fully localized phase is characterized by a Poissonian, $P(s_{\alpha}) = \exp(-s_{\alpha})$~\cite{Haake_2010}. Interestingly, we observe that the semilocalized region features  semi-Poissonian~\cite{Bogomolny_1999} statistics, $P (s_{\alpha}) = 4 s_\alpha \exp(-2s_{\alpha})$. We have checked that this behavior appears in the entire semilocalized region and is independent of $N$. The semi-Poissonian form can be simply understood for $J=0$. Then, bare ($g_c=0$) levels follow a Poisson distribution. Since for strong coupling ($g_c > W$) dark states lie in between the bare levels, we can model the dark state distribution as
\begin{equation*}
P (s_{\alpha})  = \int \! \text{d} x \text{d} y \; \delta \left(s_{\alpha} - \frac{x+y}{2}\right) {\rm e}^{-x} {\rm e}^{-y} = 4 s_{\alpha} {\rm e}^{-2s_{\alpha}},
\end{equation*}
where we assumed the hybridized states equidistant from the two closest bare levels. To check the validity of this assumption for $J\neq 0$, we analyze numerically the disorder-averaged deviation
\begin{align}
\overline{\Delta}_{\alpha}=N\left(E_{\alpha} - \frac{w_{\bi} +w_{\bi +1}}{2}\right)
\end{align} 
in Fig.~\ref{fig:fig3}\textbf{(b)}, with $w_{\bi}$ and $w_{\bi +1}$ the closest bare levels immediately below and above $E_\alpha$. While in the localized phase (triangle) the eigenenergies are found to be very close to the  bare levels, they are much closer to $(w_{\bi} +w_{\bi +1})/2$ in the semilocalized region (square), confirming our simple argument above.

\section{Dark state transport}
\label{sec3}

Finally, we discuss the role of semilocalized states on transport and diffusion. We have seen that localization properties in the semilocalized regime can be well understood for $J=0$. In this case, the dimensionality of the problem becomes irrelevant (see Appendix \ref{sec:app:retprob}). Therefore, for simplicity, we here focus on transport in 1D for a chain with sites $i = 1,\dots, N$.

We expect that generally, the semilocalized dark state can very efficiently contribute to transport. Since the polaritonic states feature homogeneous amplitudes $a_{\pm,i} \sim 1/\sqrt{N}$ throughout the system, they contribute to the infinite-time averaged transmission probability, $\Pi_{1N} \sim \sum_{\alpha} \vert a_{\alpha 1}\vert^{2} \vert a_{\alpha N} \vert^{2}$, with a term $\sim 1/N^2$. In contrast, semilocalized dark states contribute with terms $\sim 1/N$ when averaging over disorder realizations. This stems from the fact that the probability for an excitation to leave a site $i$, $1-\overline{\Pi}_{ii}\sim 0.6$ is independent of $N$. Therefore the infinite-time averaged transmission probability to any other site is $\sim (1-\overline{\Pi}_{ii}) / (N-1) \sim 0.6/(N-1)$, which explains the $\overline{\Pi}_{1N} \sim 1/N$ contribution.

\begin{figure}[tb]
    \centering
    \includegraphics[width=1\columnwidth]{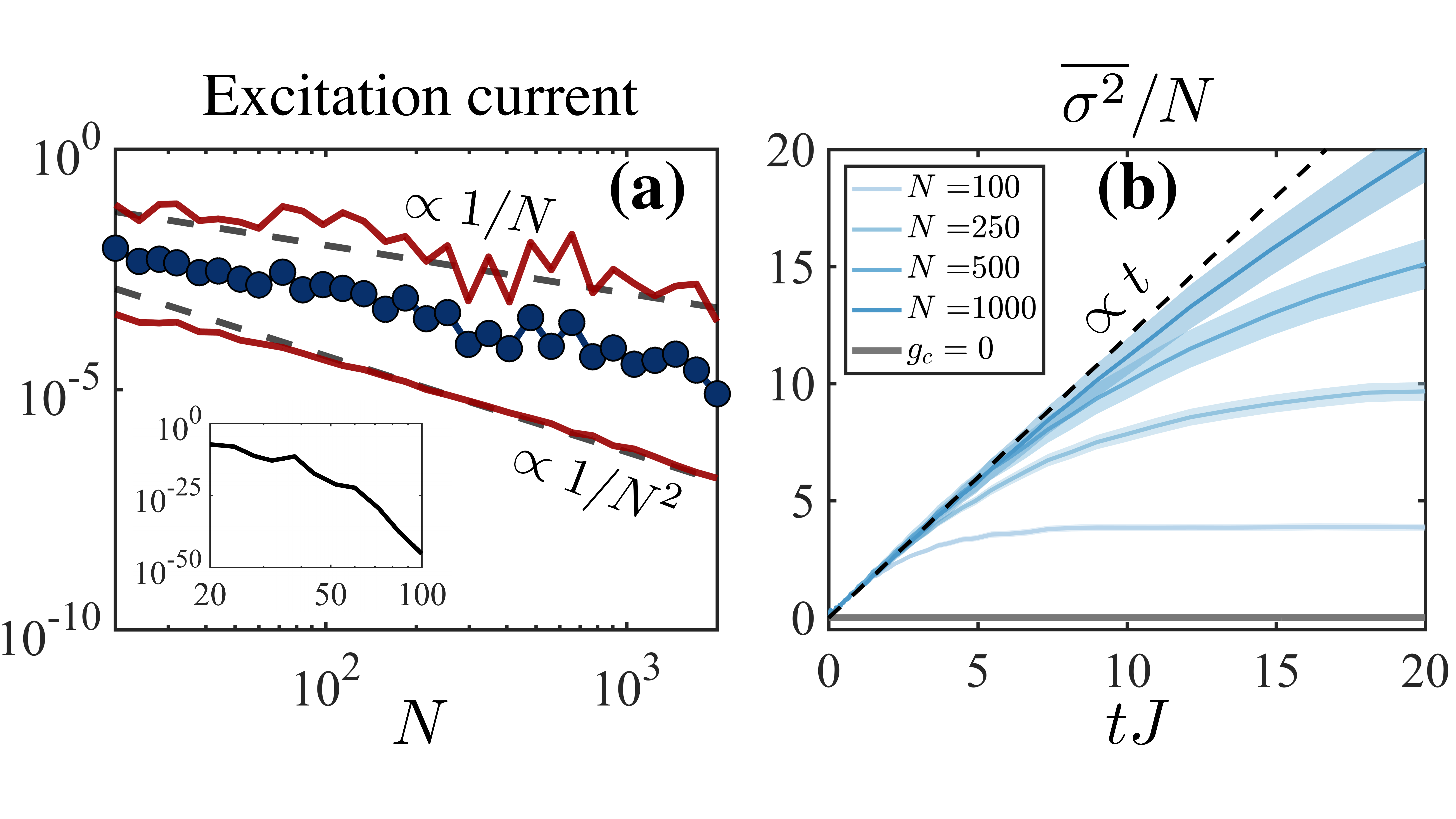}
    \caption{\textbf{(a)} Excitation currents through a 1D chain as function of $N$ (strong coupling, $g_c=30J$, $W=10J$, $\gamma = 0.05J$, currents averaged over finite times $1000 \leq tJ \leq 2000$, see text). Shown are the mean ($\overline{I}$, blue circles) and maximum/minimum currents ($I_{{\rm max}/{\rm min}}$, red lines) of $100$ disorder realizations. Dashed lines are guides to the eye for $1/N$ and $1/N^2$. Inset: $\overline{I}\propto\me^{-N}$ for $g_{c}=0$. \textbf{(b)} Disorder-averaged mean squared displacement $\overline{\sigma^{2}}/N$ (1D, $g_c = 50 J$, $W = 30 J$, $200$ realizations, shaded area: standard error of the mean, see text). While absence of diffusion is found for $g_{c}=0$ ($\overline{\sigma^{2}}/N\simeq 0$, grey line, expected for 1D Anderson localization), diffusive-like dynamics, $\overline{\sigma^{2}}\propto t$, occurs for $g_{c} \gg W$.}
    \label{fig:fig4}
\end{figure}

For a realistic transport scenario, in Fig.~\ref{fig:fig4}\textbf{(a)}, we analyze the excitation current flowing through a chain, as a function of $N$, while keeping the emitter density constant. To compute such currents, we imagine the 1D chain to be connected to two Markovian baths at the two ends. We consider the system to be initially in the state $\ket{G,0}$, turn the bath coupling on, and simulate the time-evolution. The Lindblad master equation governing the dynamics is
\begin{align}
\frac{d \hat \rho}{dt}=-\mi [\hat H,\hat \rho] + \sum_{\eta} \hat{\mathcal{L}}_{\eta} (\hat \rho),
\end{align}
with $\hat \rho$ the density matrix and two dissipative Lindblad processes adding/removing excitations on the first and last site. Here, 
\begin{align}
\hat{\mathcal{L}}_{\eta} (\hat \rho)=-\{\hat{L}_{\eta}^{\dagger} \hat{L}_{\eta},\hat \rho \} + 2\hat{L}_{\eta} \hat{\rho}\hat{L}^{\dagger}_{\eta}
\end{align} 
with $\hat{L}_{\rm in}=\sqrt{\gamma/2} \hat{\sigma}^{+}_{1}$ and $\hat{L}_{\rm out}=\sqrt{\gamma/2} \hat{\sigma}^{-}_{N}$ with $\gamma$ a pumping rate. The excitation current can be computed as $I\equiv{\rm Tr}[\hat{\sigma}^{+}_{N} \hat{\sigma}^{-}_{N} \hat \rho]$~\cite{Schachenmayer_Cavity_2015}. In contrast to previous simulations of such scenarios (e.g.~in~\cite{Schachenmayer_Cavity_2015}), there are no additional dissipative decay channels in our coherent transport model. We therefore find that the dynamics of $I$ exhibits persistent small oscillations up to long times. We therefore average over the time-scale $1000 \leq tJ \leq 2000$. At those late times, we still observe a very slow decrease of the current with time for the smaller system sizes. For $N>1000$ we don't find any significant evolution. 
The data in Fig.~\ref{fig:fig4} is additionally averaged over $100$ disorder realizations. For $g_c=0$ we always find an exponentially suppressed current, $\overline{I} \sim \exp(-N)$, while for strong coupling ($g_c = 30J$) the mean current decays slower, closer to a $\overline{I}\sim 1/N$ scaling. Additionally, we also plot maximum (minimum) currents $I_{\rm max}$ ($I_{\rm min}$) of the realizations. For large $N$, $I_{\rm min}$ decreases as $\sim 1/N^2$ and exhibits only small fluctuations. We interpret this as an ``unlucky'' disorder realization prohibiting efficient dark-state transport, requiring the energy to flow through polaritonic states~\cite{Schachenmayer_Cavity_2015}. We note that it is not clear whether the $1/N$ scaling of the finite time currents from Fig.~\ref{fig:fig4}\textbf{(a)} can be related to the $\sim 1/N$ contribution in the infinite time averaged quantity $\overline{\Pi}_{1N}$~\cite{Chavez_Disord_2020}.

In Fig.~\ref{fig:fig4}\textbf{(b)} we analyze the diffusion properties in the 1D chain, after initializing the system in the state $\ket{\phi(t=0)}=\ket{N/2,0}$. We show the time evolution of the disorder-averaged mean squared displacement $\overline{\sigma^{2}}= \sum_{j} \vert i-j \vert^{2} \overline{P}_{ij}(t)$. For $g_{c}=0$ and $W=30J$, eigenstates are fully localized and diffusion is suppressed, $\overline{\sigma^{2}}\sim \textrm{cst}$, while a diffusive-like behavior $\lim_{N\to \infty}\overline{\sigma^{2}}\propto t$ occurs in the strong coupling case ($g_c=50 J$) up to finite size effects. Second order perturbation theory  (Schrieffer-Wolff transformation~\cite{Zhu_2013}) leads to an effective {\em correlated} hopping model (see Appendix \ref{sec:app:golden} for a detailed derivation), with on-site energy dependent amplitude, which differs from other known disordered models, e.g.~with power-law hopping~\cite{Levitov1990,Mirlin1996,Celardo2016,Deng2016,Deng2018}. Diffusive-like behavior can  be qualitatively understood for $J=0$: the transition probability $P_{ij}(t)$ is then not correlated with distance $|i-j|$, so  
\begin{align}
\overline{\sigma^{2}} 
&= \sum_{j} |i-j|
^2 \sum_{k\neq i} \frac{\overline{P}_{ik} (t)}{N} \nonumber \\
&\simeq \frac{1-\overline{P}_{ii}(t)}{N} \sum_{j} |i-j|
^2  \propto N^2 \left[1-\overline{P}_{ii}(t)\right].
\end{align}
The escape probability can be estimated to increase linear in time, $1-P_{ii}(t) \propto t/N$, with Fermi's golden rule for large $N$ (as shown in Appendix \ref{sec:app:golden}). We note that this behavior does not correspond, strictly speaking, to a diffusive dynamics since the increase of the mean squared displacement is non-local and stems from the evenly distributed growth of the probability amplitudes $a_{\alpha\bj}$ over the whole chain [see Fig.~\ref{fig:fig1}\textbf{(d)}].

\section{Conclusion}
\label{sec5}

We have shown that Anderson localization can be strongly modified by coupling the disordered ensemble to a cavity. This is manifested by the emergence of dark states localized on multiple sites with energy spacings following semi-Poissonian statistics. We denote such states as {\em semilocalized}. We find that typical localization quantifiers such as the IPR exhibit properties common to ordinarily localized states (constant scaling with system size), but at values below one, $\overline{{\rm IPR}}(\epsilon_\alpha) \sim 0.4$. Additionally, in this semilocalized regime, the level-spacing statistics exhibits a semi-Poissonian behavior, which neither corresponds to ordinary localized states (Poissonian distribution) nor extended states (Wigner-Dyson distribution). We further analyzed the contribution of semilocalized states to transport, and found that they are responsible for a diffusive-like behavior and an algebraic decay of energy transmission for strong light-matter couplings. It is an interesting prospect to investigate how dissipation~\cite{Celardo_2013} affects the transport properties of such states.

\bigskip

\section*{Acknowledgements}
We are grateful to Denis Basko,  Claudiu Genes, Nikolay Prokof'ev, Antonello Scardicchio, Giuseppe Luca Celardo, Francesco Mattiotti and Fausto Borgonovi for stimulating discussions. This work was supported by the ANR - ``ERA-NET QuantERA'' - Projet ``RouTe'' (ANR-18-QUAN-0005-01), and LabEx NIE (``Nanostructures in Interaction with their Environment'') under contract ANR-11-LABX0058 NIE with funding managed by the French National Research Agency as part of the ``Investments for the future program'' (ANR-10-IDEX-0002-02), and IdEx Unistra project STEMQuS.  G. P. acknowledges support from the Institut Universitaire de France (IUF) and the University of Strasbourg Institute of Advanced Studies (USIAS). Research was carried out using computational resources of the Centre de calcul de l'Universit\'e de Strasbourg. 

\section*{Appendix}

\appendix

In the following Appendixes we provide further details on analytical calculations: We discuss our perturbation theory (Appendix \ref{sec:app:pert}), arrowhead matrix calculations (Appendix \ref{sec:app:arrow}), and the Schrieffer-Wolff transformation for estimating diffusion with Fermi's golden rule  (Appendix \ref{sec:app:golden}). We provide results for the return probability also in 1D systems (Appendix \ref{sec:app:retprob}). Lastly, we describe a scheme how one could implement our model in recent cold atom experiments (Appendix \ref{sec:app:AMO}).

\section{Perturbation theory} 
\label{sec:app:pert}
We start from the Hamiltonian $\hat H_{0}$ without hopping ($J=0$), and treat the light-matter coupling contribution $H_{\rm I}$ as a perturbation. To second order, the eigenstates
\begin{align}
\ket{\bi,0}^{(2)} =\ket{\bi,0} + \frac{g}{w_{\bi}+\delta} \ket{G,1}+ \sum_{\bj\neq \bi} b_{\bi\neq\bj} \ket{\bj,0}
\label{ffu}
\end{align}
acquire a finite amplitude on site $\bj \neq \bi$ via the cavity: 
\begin{align}
b_{\bi\neq\bj} \equiv \langle \bj,0 \ket{\bi,0}^{(2)}= \frac{g^2}{(w_{\bi}-w_{\bj})(w_{\bi} + \delta)}.
\end{align}
Note that (taking $\delta=0$), the individual perturbative state amplitudes in Eq.~(\ref{ffu}) are valid for $g^{2} \ll |w_{\bi} (w_{\bi}-w_{\bj})|$, which is satisfied in the thermodynamic limit $N\to \infty$ for a fixed $g_{c}$, as long as $w_{i}$ is not too close to the middle of the distribution $w_{i}=0$ and for $w_{\bi}$ and $w_{\bj}$ not accidentally in close resonance.

An analytic expression of the constant tail [Fig.~1\textbf{(c)} of the main text] can be derived in the perturbative regime by computing the disorder average of the squared amplitudes $b_{\bi\neq\bj}$ in the limit $N\to \infty$. For $\delta=0$, we obtain
\begin{align}
\overline{|b_{\bi\neq\bj}|^2}&=\frac{1}{N} \sum_{\bi}\sum_{\bj\neq \bi} |b_{\bi\neq\bj}|^2 \nonumber \\
&= \frac{g^{4}}{N} \int_{-W/2}^{W/2} d\omega \frac{\rho (\omega)}{\omega^{2}}\int_{-W/2}^{W/2} d\omega' \frac{\rho (\omega')}{(\omega-\omega')^{2}}, 
\label{ffu2}
\end{align}
with the uniform density of states $\rho(\omega)=N/W$. Because of the divergence for $\omega=\omega'$ occurring in the second integral, we compute the Hadamard finite part of the latter 
\begin{align}
\mathcal{H} \int_{-W/2}^{W/2} d\omega' \frac{\rho (\omega')}{(\omega-\omega')^{2}} &= \lim_{\epsilon \to 0} \Bigg[\int_{-W/2}^{\omega-\epsilon} d\omega' \frac{\rho (\omega')}{(\omega-\omega')^{2}} \nonumber \\
&+ \int_{\omega+\epsilon}^{W/2} d\omega' \frac{\rho (\omega')}{(\omega-\omega')^{2}} - \frac{2}{\epsilon}\Bigg] \nonumber \\
&=\frac{W}{\omega^{2}-(W/2)^{2}}.
\end{align}
Using this result in Eq.~(\ref{ffu2}) and considering again the finite part (removing the divergences at $\omega=0$ and $\omega=\pm W/2$) of the remaining integral, we obtain the result given in the main text:
\begin{align}
\overline{|b_{\bi\neq\bj}|^2}=\frac{4g_{c}^4[4-2\log(4)]}{N W^{4}} \qquad (g_{c}=g\sqrt{N}).
\end{align}
This prediction agrees very well with numerically exact simulations (see main text) for $g_c \ll W$, indicating that occasional resonances and the divergence at $\omega=0$ do not play an important role for the averaged tails of the eigenstates.

\section{Arrowhead Hamiltonian} 
\label{sec:app:arrow}

In the single-excitation subspace, the Hamiltonian $\hat H$ without hopping ($J=0$) takes the form of an ``arrowhead'' matrix in the basis $\{\ket{\bi,0}$, $\ket{G,1}\}$:
\begin{align}
	\hat H = \left( \begin{array}{cccccc|c}
		w_1   & 0 & & & \dots & 0 & g \\
		0   & w_2 &  0 &  & &  &  \\
		     &  0 & \ddots  & \ddots & &  \vdots &  \\
		\vdots     &  & \ddots   &  & 0 &  & \vdots  \\
		     &  &   &  0 & w_{N-1} & 0 & \\
		0   &  & \dots  &  & 0 & w_N  & g \\ \hline
		g & &  & \dots && g  & 0
	\end{array} \right).
\end{align}
A direct property of this arrowhead form is that after sorting the bare energies $w_{\bj}$ in increasing order, i.e. $w_1 \leq w_2 \leq \dots \leq w_N$, the eigenvalues of $H$ are interlaced with those bare energies~\cite{OLEARY_1990}:
\begin{align}
	\epsilon_{-} \leq w_1 \leq  \epsilon_1 \leq w_2 \leq \dots  \leq w_{N-1} \leq \epsilon_{N-1} \leq w_N \leq \epsilon_{+}.
\end{align}
In the strong coupling case, the two eigenvalues $\epsilon_{\pm}$ at the edges of the spectrum correspond to the polariton state frequencies $\sim \pm g_c$, while the remaining ones $\epsilon_{1},\epsilon_{2},\cdots,\epsilon_{N-1}$ correspond to the $N-1$ dark state frequencies denoted as $\epsilon_{\alpha}$ in the main text.
The $N+1$ eigenstates satisfying $\hat{H} \psi_\alpha = \epsilon_\alpha \psi_\alpha$ take the form~\cite{OLEARY_1990}
\begin{equation}
	\label{eq:eigenstates}
	\psi_\alpha=  \frac{1}{\sqrt{1 + \frac{1}{N} \sum_{j=1}^N \frac{g_c^2}{(\epsilon_\alpha - w_j)^2} }} \left( \begin{array}{c}
		\frac{g_c/\sqrt{N}}{\epsilon_\alpha - w_1} \\
		\frac{g_c/\sqrt{N}}{\epsilon_\alpha - w_2} \\
		\vdots \\
		\frac{g_c/\sqrt{N}}{\epsilon_\alpha - w_N} \\
		1
	\end{array} \right),
\end{equation}
with $\alpha= \pm , 1,\dots , N-1$. The photon weight is the squared amplitude of the $(N+1)^{\rm th}$ component,
\begin{equation}
    {\rm PW}_\alpha  = \left| \psi_{a, N+1} \right|^2 =  \frac{1}{1 + \frac{1}{N} \sum_{j=1}^N \frac{g_c^2}{(\varepsilon_\alpha - w_j)^2} } .
\end{equation}
For a dark eigenstate $\psi_\alpha$ ($a= 1, \dots , N-1$), the sum in the denominator is dominated by the terms where $w_j$ is the closest to $\epsilon_\alpha$. Since the spacing between energies is of order $O(1/N)$, those terms are of order $O(N^2)$. With the factor $1/N$ in front of the sum, this implies that the denominator is of order $O(N)$. Thus, the photon weight of the dark states scales as
\begin{align}
    {\rm PW}_\alpha \, =\, O(1/N). 
\end{align}
This implies that the dark eigenstates have vanishing photon weight in the thermodynamic limit. This is in stark contrast with the photon weights of the two polariton eigenstates with energies $\epsilon_\pm$, which is of order $O(1)$. Note that this argument is generally valid,  beyond the perturbative limit.

\medskip

The arrowhead shape of the Hamiltonian can also help to better understand the diffusion properties of the system. In particular, here we show how it can be used to argue for diffusive-like behavior of the mean squared displacement of an excitation, $\overline{\sigma^{2}} \propto t$ in the strong coupling regime and that this property originates from the contribution of dark states.  We briefly sketch the argument here; a more detailed treatment will be given in Ref.~\cite{dubail2020}.

\smallskip

In the absence of hopping and for random bare energies $w_i$, there is no correlation between the position of the emitters and their energy. Therefore, the probability that an excitation at emitter $a$ arrives at emitter $b$ at time $t$ depends only on their bare energies $w_a$ and $w_b$, but not on the distance between them (the dimensionality is also irrelevant). Then, upon averaging over disorder, all pairs of indices ($a,b$) with $a\neq b$ contribute the same amount to the mean squared displacement. Therefore the mean squared displacement $\overline{\sigma^2}$ is proportional to the quantity
\begin{equation}
    \overline{Q(t)} = \frac{1}{N} \sum_{1 \leq a \neq b \leq N} \overline{\left|[\me^{-\mi \hat H t}]_{ab}\right|^2},
\end{equation}
where $[\me^{-\mi \hat H t}]_{ab}$ is the $(a,b)$ entry of the $(N+1)\times (N+1)$ matrix $\me^{-\mi \hat H t}$. $\overline{Q(t)}$ measures the probability that the excitation has moved at time $t$, regardless of its initial position. Using the fact that the matrix $\me^{-\mi \hat H t}$ is unitary, one can rewrite the above sum as
\begin{align}
    \overline{Q(t)} &= \frac{1}{N} \sum_{a=1}^N (1-\overline{P_{aa}(t)}) + \frac{1}{N} (1- \overline{\left| [\me^{-\mi \hat H t}]_{N+1,N+1}\right|^2}) \nonumber \\
    & - \frac{2}{N} \sum_{a=1}^N \overline{\left| [\me^{-\mi \hat H t}]_{a,N+1} \right|^2},
    \label{eq:app:Qt}
\end{align}
where $P_{aa}(t) = \left| [\me^{-\mi \hat H t}]_{aa} \right|^2$ is the return probability of an excitation initially located on the emitter $a$. The second and third terms in (\ref{eq:app:Qt}) remain bounded as time increases, while the first term keeps increasing and quickly becomes dominant:
\begin{equation}
    \overline{Q(t)} \simeq \frac{1}{N} \sum_{a=1}^N (1-\overline{P_{aa}(t)} ).
\end{equation}
The time-dependent escape probability, $(1-\overline{P_{aa}(t)} )$, thus determines the evolution of the mean squared displacement. For large $N$, we can compute the evolution in the dark-state energy range in time-dependent perturbation theory and treat the evolution after initializing the system with an excitation on a single site $\bi$, by using Fermi's golden rule (as shown in the next Appendix Sec.~\ref{sec:app:golden}). Then, $(1-\overline{P_{aa}(t)} ) = \Gamma_a t$, with a rate $\Gamma_a$ given by Eq. (\ref{eq:app:escape}). This gives the linear growth:
\begin{align}
    \overline{\sigma^2} \propto \overline{Q(t)} \simeq \left( \frac{1}{N} \sum_{a=1}^N \Gamma_a\right) t.
\end{align}
In contrast, the two polaritonic states lead to $\overline{\sigma^{2}}\sim t^{4}$ at short times ($t\lesssim 1/g_{c}$) and generate small oscillations that are superimposed with the general linear growth predicted from the dark states~\cite{phdthesis_botzung}.

\section{Schrieffer-Wolff transformation and Fermi's golden rule}
\label{sec:app:golden}
Starting with an excitation localized on site $\bi$ at time $t=0$, we compute the escape probability $1-P_{\bi\bi}(t)$ to other sites $\bj\neq\bi$ at time $t$. From energy conservation, one can already expect that these processes imply $w_{\bi}=w_{\bj}$, and therefore the perturbative expansion Eq.~(\ref{ffu}) does not appear to be well suited as the third term in the right-hand-side diverges. It is instead convenient to use a Schrieffer-Wolff transformation for the Hamiltonian $\hat H=\hat H_{0}+\hat H_{\rm I}$, which results in a disentanglement of light and matter degrees of freedom~\cite{Zhu_2013}. The new Hamiltonian is written as $\hat H'=\me^{\hat S}\hat H \me^{-\hat S}$. Under the assumption that the eigenvalues of the generator $\hat S$ remain small (see below), one can expand $\hat H'$ as $\hat H'=\hat H+[\hat S,\hat H]+\frac{1}{2}[\hat S,[\hat S,\hat H]]+ \cdots$. The linear coupling term $\hat H_{\rm I}$ can be removed from the expansion with the choice 
\begin{align}
\hat S = \sum_{i} \frac{g}{w_{\bi}+\delta} (\hat a \hat \sigma^{+}_{\bi} - \hat \sigma^{-}_{\bi} \hat a^{\dagger}),
\end{align}
which provides $[\hat S,\hat H_{0}]=-\hat H_{\rm I}$. 
The new effective Hamiltonian (for $J=0$) then takes the form  
\begin{align}
\hat H' = \hat H_{0}+\frac{1}{2} [\hat S, \hat H_{\rm I}] + \mathcal{O}(\hat H_{\rm I}^{3}),
\label{ffu5}
\end{align}
and the condition to be satisfied if one is to keep only the first two terms on the right-hand-side of Eq.~(\ref{ffu5}) is $g \ll |w_{\bi}+\delta|$. Calculating the commutator $[\hat S, \hat H_{\rm I}]$, we obtain
\begin{align}
\hat H' &= 
\omega_{c}\hat{a}^\dag \hat{a} 
+ \sum_{\bi} (\omega_e + w_{\bi}) \hat \sigma^{+}_{\bi} \hat \sigma^{-}_{\bi} + \hat a^{\dagger} \hat a \sum_{\bi} \frac{4g^{2}}{w_{\bi}+\delta} \hat \sigma^{+}_{\bi} \hat \sigma^{-}_{\bi} \nonumber \\
& + \frac{g^{2}}{2}\sum_{\bi,\bj} \left(\frac{1}{w_{\bi}+\delta} + \frac{1}{w_{\bj}+\delta} \right) \hat \sigma^{+}_{\bi} \hat \sigma^{-}_{\bj},  
\label{ffu6}
\end{align}
up to a constant term. The last term corresponds to an effective correlated hopping between arbitrarily distant sites, while the third term results in a renormalization of the cavity frequency depending on the two-level emitter states. This term does not contribute to transitions between states with one excited emitter and zero photon, and can therefore be dropped out of the calculation. Because of the absence of divergence for $w_{\bi}=w_{\bj}$, the Hamiltonian Eq.~(\ref{ffu6}) is well suited to compute the escape probability from site $\bi$ using Fermi's golden rule. The latter reads $1-P_{\bi\bi}(t)=\Gamma_\bi t$, where the escape rate is (for constant $\rho(\omega) = N/W$):
\begin{align}
\Gamma_\bi &=2\pi \int_{-W/2}^{W/2} d\omega \rho(\omega) \vert\langle \bj,0 \vert \hat V \vert \bi,0 \rangle \vert^{2} \delta (\omega-w_{\bi}) \nonumber \\
&=\frac{2\pi g^{4}_{c}}{N W (w_{\bi}+\delta)^2}.
\label{eq:app:escape}
\end{align}
Here, $\hat V$ corresponds to the second term in Eq.~(\ref{ffu6}) with matrix elements (written as function of the continuous variable $\omega$)
\begin{align}\langle \bj,0 \vert \hat V \vert \bi,0 \rangle =\frac{g^{2}}{2} \left(\frac{1}{w_{\bi}+\delta} + \frac{1}{\omega +\delta} \right).
\end{align}
Note that the two conditions $1/W \ll t \ll N/W$ and $1-P_{\bi\bi}(t) \ll 1$ ensuring validity of the Fermi golden rule can be satisfied simultaneously in the thermodynamic limit $N\to \infty$. A lower bound for the (normalized) mean squared displacement can be estimated from Eq.~(\ref{eq:app:escape}):
\begin{align}
    \label{eq:app:escape2}
\frac{\overline{\sigma^2}}{N} =\frac{1}{N} \sum_{\bi,\bj} \frac{\vert \bi-\bj \vert^{2}}{N^{2}} \times \frac{1}{N} \sum_{\bi} \Gamma_{\bi} t
\geq  
\frac{\pi g_c^4 t}{3 W(W/2 + |\delta|)^2}.
\end{align}
This Fermi's golden rule result overestimates the numerical data plotted in Fig.~4\textbf{(b)} of the main text. However, those numerical results can be alternatively well described analytically using an exact formula derived directly from Eq.~(\ref{eq:app:Qt})~\cite{dubail2020}.

\section{Return probability in 1D}
\label{sec:app:retprob}

\begin{figure}[t]
    \centering
    \includegraphics[width=0.5\textwidth]{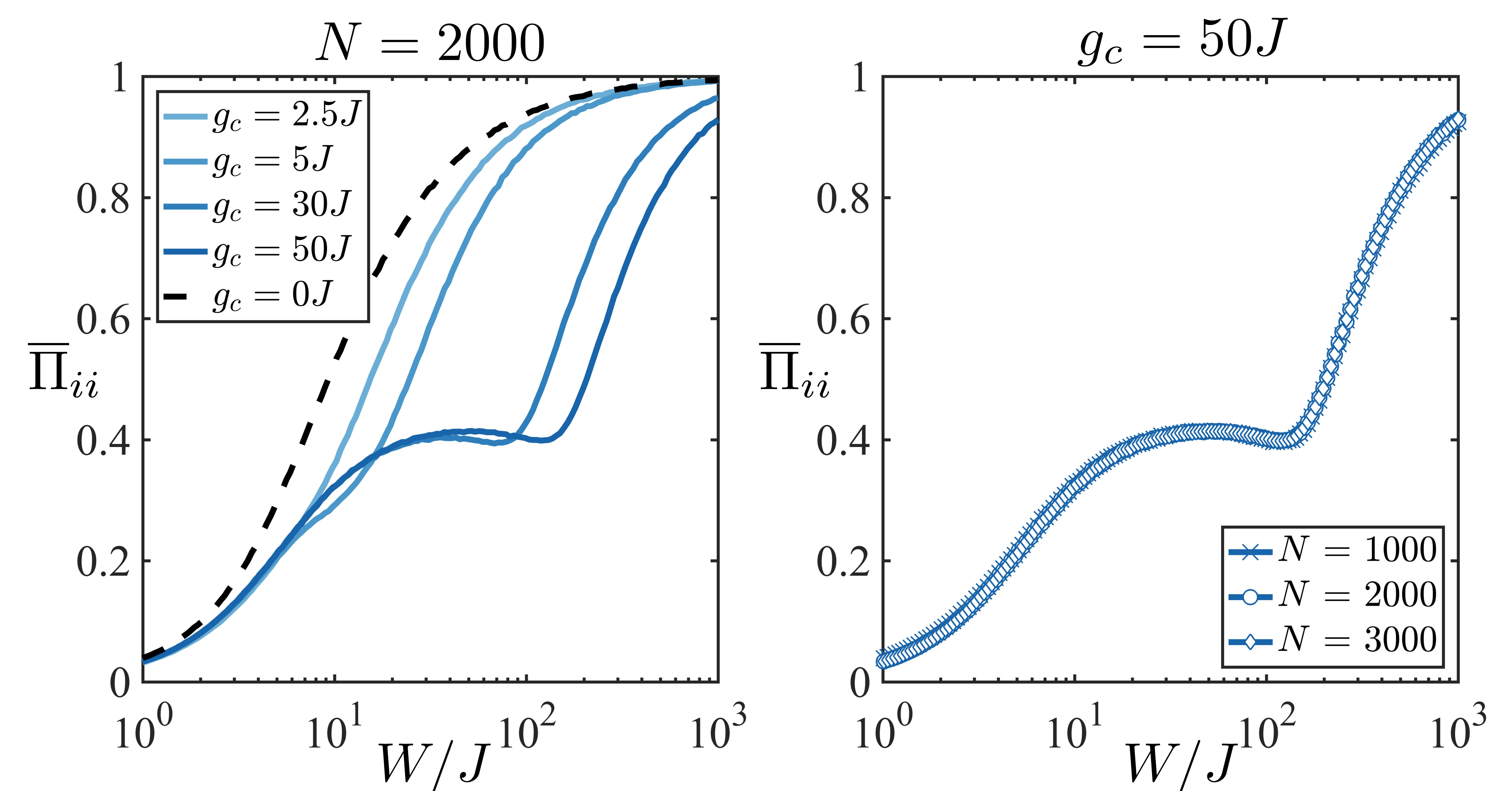}
    \caption{\textbf{(a)} Disorder-averaged return probability $\overline{\Pi}_{\bi \bi}$  as a function of $W/J$ (for a central site $\bi$ of a $N = 2000$ chain, mean emitter splitting on resonance with cavity, $\delta=0$). For strong-couplings $g_c > W > W_{c}$, a plateau ($\overline{\Pi}_{\bi \bi}\simeq 0.4$) indicates the ``semilocalized'' regime. \textbf{(b)}  $\overline{\Pi}_{\bi \bi}$  as a function of $W/J$ for $g_c=50$ and various system sizes $N$, demonstrating independence on $N$.}
    \label{fig:ret_prob_1d}
\end{figure}

In Fig.~2(a) of the main paper we observed the appearance of the semilocalized regime by the presence of a plateau in the return-probability, which is independent of $N$. Here, we verify that this physics is indeed independent on the dimensionality, see Fig.~\ref{fig:ret_prob_1d}.

\pagebreak

\section{Model realization with cold atoms}
\label{sec:app:AMO}

Here we discuss possible experimental implementations to observe the physics of semilocalized states with cold atoms. Such setups have been recently used to study coherent physics of the Dicke model without disorder in various scenarios. 

\begin{figure}[t]
    \centering
    \includegraphics[width=0.29\textwidth]{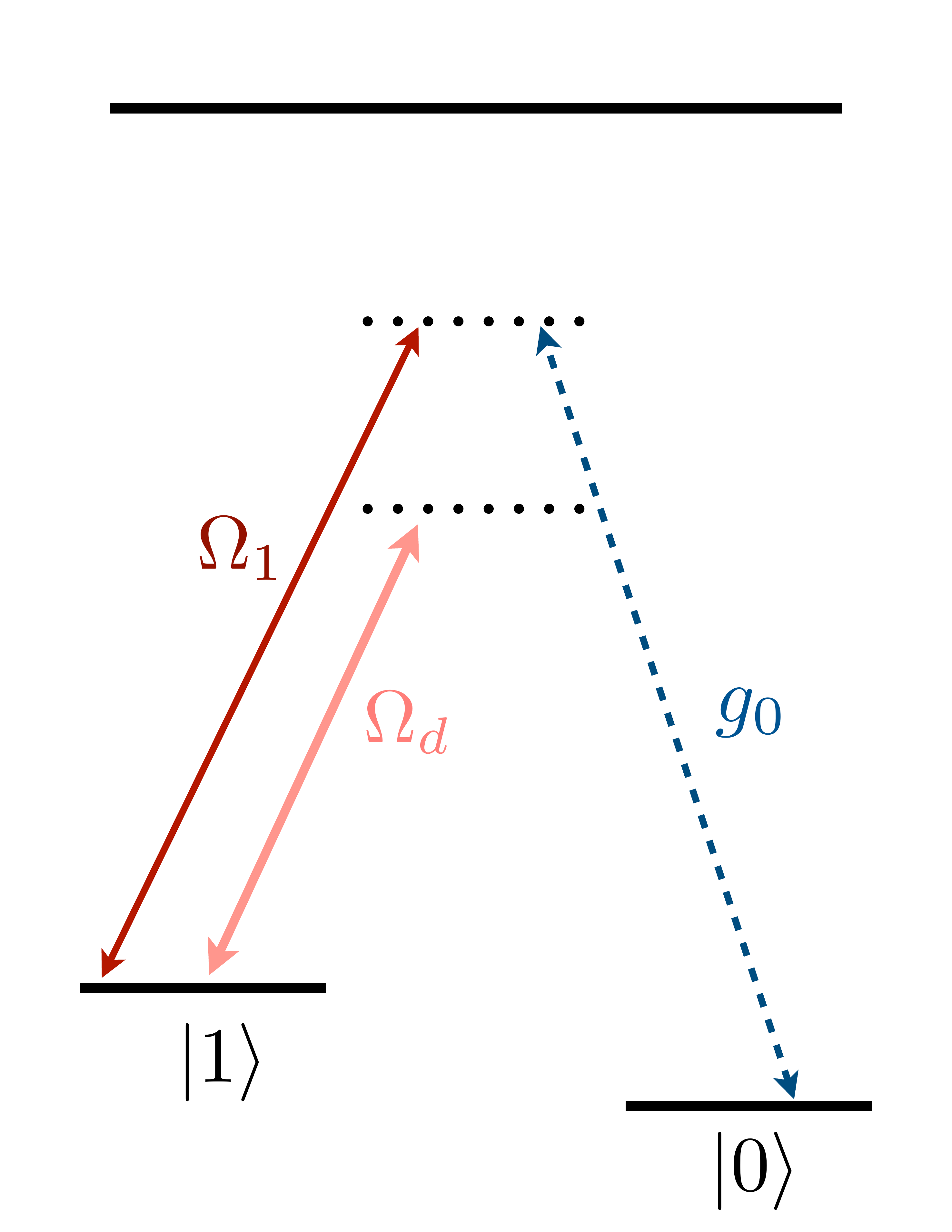}
    \caption{Proposed level scheme for engineering the disorder Tavis-Cummings model with cold atoms. The system is based on the Raman dressing scheme from Ref.~\cite{Dimer_Propo_2007} that has led to experimental observation of Dicke model physics in~\cite{Zhang_Dicke_2018}.}
    \label{fig:level_scheme}
\end{figure}

To avoid decay of atomic excited states, it has been proposed to realize the Dicke Hamiltonian with two internal hyper-fine ground states of, e.g., alkali atoms, as effective emitter states. This can be achieved by using a Raman laser dressing for atoms trapped within an optical cavity, as first suggested in \cite{Dimer_Propo_2007}. Here, each atom has two long-lived ground states $\ket{0}$ and $\ket{1}$ that are coupled to the cavity mode (coupling strengths $g_{0}$ and $g_1$), and through a pair of balanced lasers, de-tuned from an excited state, with respective Rabi frequencies $\Omega_0$ and $\Omega_1$ (de-tunings $\Delta_0$ and $\Delta_1$). After adiabatically eliminating the atomic excited state, the two laser couplings lead effectively  to resonant and counter-rotating Dicke model terms, with individually tunable Hamiltonian contributions~\cite{Dimer_Propo_2007}:
\begin{align}
    \hat H \propto  \frac{g_0\Omega_1}{2\Delta_1} \sum_{\bi} (\hat a \hat \sigma_{\bi}^{+} + \hat a^\dag \hat \sigma_{\bi}^{-} )
    +
    \frac{g_1\Omega_0}{2\Delta_0} \sum_{\bi} (\hat a^\dag \hat \sigma_{\bi}^{+} + \hat a \hat \sigma_{\bi}^{-} )
\end{align}
To realize our Tavis-Cumming model, one can remove one laser, $\Omega_0 = 0$, see Fig.~\ref{fig:level_scheme} for a sketch of the level scheme. This proposal has been recently experimentally realized in~\cite{Zhang_Dicke_2018}, where the two hyper-fine ground states $\ket{F= 1, m = 1} \equiv \ket{0}$ and $\ket{F= 2, m = 2} \equiv \ket{1}$ of $^{87}$Rb have been used. This particular experiment used a cavity with decay rate $\kappa = 2\pi \times 0.1$~MHz. Atoms are trapped in an intra-cavity optical lattice ensuring a location of the atoms at cavity anti-nodes. Such schemes allow to trap and collectively couple $\sim {10^5}$ atoms~\cite{Zhiqiang2017}, and can lead to effective collective coupling strengths of $g_c = \frac{g_0\Omega_1}{2\Delta_1} \sqrt{N} \gtrsim \kappa$ overcoming the cavity decay rate. Reachable coupling strengths are simply a problem of achievable atom numbers and laser powers. Note that as an alternative to atomic ground states, also long-lived excited states of alkaline earth atoms may also be used. For example, in experiments as in~\cite{Muniz2020}, $^1$S$_0$ and $^1$P$_1$ states of Strontium atoms have been successfully used as effective emitter states to engineer infinite-range spin-models (after adiabatically eliminating the cavity).

Adding disorder to this setup can  be readily achieved by superimposing additional external light fields that selectively (via the polarization) induce AC Stark shifts to one of the ground state levels (see additional disorder laser field with Rabi frequency $\Omega_d$ in Fig.~\ref{fig:level_scheme}). For example, if atoms are trapped in a regular 3D optical lattice, this can e.g.~be achieved with a second incommensurate lattice of different wavelength~\cite{Roati_Ander_2008,Schreiber_Obser_2015}. In a 1D optical lattice, random positions of atoms within ``pancakes'' would naturally lead to disorder. The intensity of this additional disorder light field can be sufficiently small, since the required disorder strengths for typical experimental numbers of~\cite{Zhang_Dicke_2018,Zhiqiang2017} correspond to only $W \approx 100 \,{\rm Hz} \ll g_c$ and are thus not hampering with the Raman dressing scheme. Even naturally existing disorder due to stray fields may be beneficial to study the physics described in our main text.

A cold atom setup as in ~\cite{Zhang_Dicke_2018,Zhiqiang2017} could be used for observing excitation diffusion as proposed in Fig.~4\textbf{(b)} in the main text, after selectively creating population in one of the two emitter states of some atoms. Diffusion between ``pancakes'' in 1D optical lattices could be directly monitored or accessed via time-dependent measurements of emitter state populations. One could also homogeneously couple 1000s of atoms with a 3D intra-cavity optical lattice in order to study diffusion on a regular lattice. This can be achieved by mode-matching the lattice lasers with a cavity field as described in~\cite{Wellnitz_Colle_2020}. 

Finally, we remark that the diffusive-like dynamics (see e.g.~Fig.~4\textbf{(b)} in the main text) can occur on very fast time-scales, as the mean squared displacement is proportional to the emitter number  $\overline{\sigma^2} \propto N$. With experimentally achievable emitter numbers the diffusion rates could be made large enough such that trapping does not play a crucial role during diffusion. It may then also be practical to use (un-trapped) atomic excited state as emitter states directly. Then, models with $J>0$ can also be considered, using atomic excitation hopping models, e.g.~with Rydberg atoms as proposed in~\cite{Schachenmayer_Cavity_2015}. Systems using rotational states of intra-cavity trapped polar molecules may also be an alternative~\cite{Schachenmayer_Cavity_2015}.

\pagebreak

\bibliography{main}

\end{document}